\documentstyle[prc,preprint,aps]{revtex}
\begin{document}
\draft
\title{Mapping the proton drip-line up to $A=70$}
\author{W.~E. Ormand}
\address{Dipartimento di Fisica, Universit\`a Degli Studi di Milano, and \\
Istituto Nazionale di Fisica Nucleare, Sezione di Milano,\\
Via Celoria 16, 20133 Milano, Italy,\\
Physics Division, Oak Ridge National Laboratory, P.O. Box 2008, \\
MS-6373 Building 6003, Oak Ridge, TN 37831-6373\\
and\\
Department of Physics and Astronomy, 202 Nicholson Hall,\\
Louisiana State University, Baton Rouge, LA 70803-4001}
\maketitle
\begin{abstract}
Coulomb energy differences between mirror nuclei with $A \le 70$ are 
calculated within the framework of the nuclear shell model using an
effective Coulomb plus isovector and isotensor interaction. Absolute
binding energies for proton-rich nuclei are predicted by adding the
calculated Coulomb shifts to experimentally measured binding energies 
for the neutron-rich mirror. The location of the proton drip-line is
investigated, as well as candidates for the exotic decay mode
known as di-proton emission. Taking into account the lifetimes of 
competing decay modes and limits imposed by experimental setups, it is
concluded that the best candidates for the observation of correlated 
di-proton emission are $^{45}$Fe, $^{48}$Ni, and $^{63}$Se. 
\end{abstract}

\section{Introduction}
\label{s:intro}

The structure of exotic nuclei, i.e., nuclei with extreme isospin values,
is one of the most exciting challenges in low-energy nuclear physics
today. Detailed theoretical studies of exotic nuclei, when confronted
with experiments, will yield important information about the interaction 
between nucleons in the nucleus and the validity of our models for the 
structure of nuclei. In addition, the study of exotic nuclei is
essential to many fundamental issues in physics today; in particular, 
the weak interaction and nuclear astrophysics. For example, it
is believed that many of the heavy elements in the universe are produced
by the radiative capture of neutrons ({\it r}-process)~\cite{r:Bur57} or 
protons ({\it rp}-process)~\cite{r:Wal81} on unstable nuclei. 
The competition between beta decay
and particle capture traces out a path that synthesizes the known
elements. The details of this path, and, hence, the abundance of the
elements produced, depends on the temperature of the site, as well as
explicit nuclear properties such as binding energies, level densities,
spectroscopic factors, and beta-decay lifetimes. 

An additional new feature of proton-rich nuclei that will be explored in the 
next few years is the possibility of a new decay
mode known as diproton emission. Because of the pairing interaction, a 
nucleus with an even number of protons $(Z,N)$ is generally more tightly 
bound than a $(Z-1,N)$ nucleus, but, because of the symmetry energy and 
Coulomb repulsion, it may be unbound relative to the $(Z-2,N)$ system.
The number of candidates for the observation of this decay mode, however,
is sharply limited by the two-proton separation energy. This is in part due
to the fact that $\beta^+$ emission is a competing
decay mechanism, and, because of the large Coulomb energy difference between 
the parent and daughter nuclei, the 
beta-decay lifetimes are of the order 1-100~ms. In addition, a
further constraint on the observation of diproton emission 
can be imposed by the experimental apparatus, 
since, in many experiments, the parent nucleus must live long enough to be
identified. Generally speaking, these two practical constraints limit the 
observable lifetime for diproton decay to $10^{-8}-10^{-3}$~s.
On the other hand, the decay rate for diproton emission is determined by
the probability to penetrate through the Coulomb barrier, which, in
turn, is exponentially dependent on the two-proton separation energy. 
As will be shown here, the number of candidates for which the 
observation of diproton decay is practical, is limited to nuclei with
two-proton separation energies between 0.9 and 1.4~MeV. 

One of the principal motivations for the construction of radioactive beam 
facilities is to study the properties of nuclei near the limits of 
stability. Very few nuclei near the proton drip-line have been
identified, and the heaviest and most proton-rich nucleus observed to date 
is $^{49}$Ni~\cite{r:Bla96}. Even more difficult than the identification of 
an exotic nucleus is the measurement of its mass, and, at present, 
predictions regarding the {\it r}- and {\it rp}-process, must rely 
on theoretical estimates for nuclear binding energies.

Several methods have been used to obtain theoretical estimates for 
absolute binding energies. One is the liquid-drop formula and associated 
variants, such as the microscopic-macroscopic approach~\cite{r:macro,r:Mol88}. 
In general, these models are determined by fitting a set of  
liquid-drop parameters while including effects due to pairing and shell 
corrections to experimental data over a wide range of nuclei, and have 
been found to reproduce known nuclear masses at the level of approximately 
800~keV (see for example Ref.~\cite{r:Mol88}). Although the
microscopic-macroscopic approach gives a good global description of 
nuclear binding energies and is the method of choice for heavy
nuclei where detailed microscopic calculations are not feasible, there
are notable discrepancies between experimental and calculated binding 
energies with neutron number between 20 and 40, as is illustrated in
Fig. 1 of Ref.~\cite{r:Mol88}.

For lighter nuclei, however, more accurate binding energies
can be achieved using the nuclear shell model, since, in many cases,
it is necessary only to compute the Coulomb energy difference between
mirror nuclei~\cite{r:Bro91,r:Orm96,r:Col96}. In this paper, 
Coulomb energy differences are computed for mirror nuclei in the 
{\it fp} shell for $46\le A \le 70$. By then making use of experimental 
data for the neutron-rich members tabulated in Ref.~\cite{r:Aud93}, 
absolute binding energies 
are predicted with an estimated accuracy at the level of $50-200$~keV.
With these binding energies, two-proton separation energies are
computed and rough estimates for the lifetimes for diproton decay
are made. Given the practical constraints on the decay halflives 
mentioned above, it is found that the best candidates for the 
experimental observation of correlated two-proton
emission are $^{45}$Fe, $^{48}$Ni, and $^{63}$Se.

This paper is organized into five sections. In Section~\ref{sec:Coul}, the 
systematics of Coulomb energy differences between analog nuclei are 
discussed, while in Section~\ref{sec:Shell} a shell-model description of 
these energy shifts is presented. Candidates for the exotic decay mode known
as di-proton emission are presented and analyzed in 
Section~\ref{sec:diproton}, and concluding remarks are 
collected in Section~\ref{sec:Conclude}.

\section{Systematics of Coulomb Energy Differences}
\label{sec:Coul}

If the nuclear Hamiltonian is composed of only one- and two-body parts,
quite generally, it may be separated into three components. The
dominant part, which is also responsible for most of the nuclear binding 
energy, is due to the strong interaction and is isoscalar in nature.
The other two components are due to both the Coulomb interaction and charge 
non-symmetric parts of the nucleon-nucleon interaction, and are
isovector and isotensor in character. If the isovector and isotensor
components are weak relative to the isoscalar component, then the 
binding energies for the members of an isospin multiplet may be obtained
within the context of the isobaric mass multiplet equation 
(IMME)~\cite{r:Wig57,r:Ben79,r:foot0}
\begin{equation}
BE(A,T,T_z,i) = a(A,T,i)+b(A,T,i)T_z+c(A,T,i)T_z^2,
\label{e:IMME}
\end{equation}
where $T$ and $T_z=(Z-N)/2$ denote the isospin and its third component for 
the members of the isospin multiplet, 
$Z$, $N$, and $A=Z+N$ are the number of protons, neutrons, and 
nucleons, respectively. The label $i$ in Eq.~(\ref{e:IMME}) represents all 
other quantum numbers needed to denote the state, 
such as angular momentum, state number, etc. The coefficients $a$, $b$,
and $c$ separately depend on the isoscalar~\cite{r:foot1}, isovector, and 
isotensor components of the nuclear Hamiltonian, respectively. 

In shell-model calculations, the isoscalar part of the nuclear Hamiltonian is 
usually determined empirically by fitting to experimental binding 
energies and levels that have had the Coulomb energy subtracted off in 
an average way~(c.f. \cite{r:Wil84,r:War90,r:Ric90}). The predictive
power of these effective Hamiltonians is indicated by the rms deviation
between experimental data and calculated binding energies,
and, to date, the best empirically determined Hamiltonian is that due to 
Wildenthal~\cite{r:Wil84} for use in the $0d_{5/2}$, $0d_{3/2}$, and 
$1s_{1/2}$ orbitals, where the rms deviation between theory and 
experiment is of the order 200~keV. For the most part, this interaction 
may be thought of as indicative of the best accuracy that may be achieved 
within the framework of the nuclear shell model. For other model spaces,
such as the {\it fp}-shell (defined by the $0f_{7/2}$, $0f_{5/2}$, 
$1p_{3/2}$, and $1p_{1/2}$ orbitals), the effective shell-model Hamiltonian
is less well determined and the deviation between theory and 
experiment is somewhat larger and is of the order 300~keV~\cite{r:Ric90}. 
With this in mind, we must conclude that any attempt to compute absolute 
binding energies from first principles in the shell model would include an
uncertainty of at least 200-300~keV in the $a$ coefficient of
Eq.~(\ref{e:IMME}).

From Eq.~(\ref{e:IMME}), the binding energy difference between iosbaric
analogs with $T_z=\pm T$ is given by
\begin{equation}
BE(A,T,T_z=T,i)-BE(A,T,T_z=-T,i)=2b(A,T,i)T.
\label{e:bcoef}
\end{equation}
Therefore, the most accurate way to predict absolute binding 
energies for proton-rich nuclei whose analog has an experimentally
measured mass is to compute the $b$-coefficient for the multiplet 
(or the Coulomb energy difference) and add $2bT$
to the experimental binding energy, $BE_{exp}(A,T,T_z=-T,i)$, of the 
neutron-rich analog. The overall uncertainty in the predicted binding
energy is then of the order
\begin{equation}
\delta BE(A,T,T_z=T,i)=\sqrt{(2\delta bT)^2+
\delta( BE_{exp}(A,T,T_z=-T,i)})^2,
\label{e:diff}
\end{equation} 
where $\delta b$ is the uncertainty in the $b$-coefficient and
$\delta BE_{exp}(A,T,T_z=-T,i)$ is the uncertainty in the experimental
binding energy. In many cases, it is possible to estimate the $b$-coefficient 
with an uncertainty of the order 30-40~keV~\cite{r:Orm89}, and, therefore, it
may be possible to predict the binding energies of extreme proton-rich nuclei 
at the level of 100-200~keV.

In Refs.~\cite{r:Bro91,r:Orm96,r:Col96}, the procedure outlined
above was used to predict the absolute binding energies of nuclei with
$36 \le A \le 55$. Those three works approached Eq.~(\ref{e:diff}) 
using slightly different methods, but with about the same level of accuracy, 
as is indicated by the
overall agreement between them. In Ref.~\cite{r:Bro91}, Eq.~(\ref{e:diff}) was
evaluated using shell-model calculations for the pure $0f_{7/2}$-shell nuclei
and a weak coupling approximation for those nuclei that spanned both the
$0d_{3/2}$ and $0f_{7/2}$ orbits. In Ref.~\cite{r:Orm96}, all the Coulomb
energy differences were evaluated within the framework of the shell model
using the $0d_{3/2}$ and $0f_{7/2}$ orbits and an empirical 
isospin-nonconserving (INC) interaction~\cite{r:Orm89}.
Finally, in Ref.~\cite{r:Col96}, Eq.~(\ref{e:diff}) was  
evaluated using a method based on a parameterization of the Coulomb
displacement energies~\cite{r:Col85}. The overall success  
of these works, and agreement between them, is essentially due to their 
empirical foundations. In each, a set of parameters was fit to experimental
data, and the models were then extrapolated to predict the masses of unknown
nuclei. This work is an extension of Ref.~\cite{r:Orm96} in which 
absolute binding energies of proton-rich nuclei with $46\le A \le 70$ 
are predicted by computing the Coulomb displacements within the framework of 
the nuclear shell model. 

Before continuing with the details and the results of the shell-model
calculations, it is instructive to examine the systematic behavior of the 
Coulomb displacement energies. Indeed, one of the reasons for the success of 
the three different methods is the smooth behavior as a function of nucleon
number $A$ exhibited by experimental $b$-coefficients. In addition, for a given 
mass number, the $b$-coefficients are essentially constant to within 100~keV 
or so, as can be seen from Tables 3-7 in Ref.~\cite{r:Orm89}. This behavior is 
easily understood from the liquid-drop model, where the Coulomb energy 
of a sphere
of radius $R=r_0A^{1/3}$ with charge $Ze$ is given by
\begin{equation}
E_C=\frac{3}{5}\frac{(Ze)^2}{R}.
\end{equation}
The Coulomb energy difference between analog nuclei is then
\begin{eqnarray}
\Delta E_C &=& 
\frac{3}{5} \frac{e^2}{R}[Z^2-(A-Z)^2] =
\frac{3}{5} \frac{e^2}{R}A(Z-N)\\
&=&\frac{3}{5} \frac{e^2}{R}2AT = \frac{3}{5} \frac{e^2}{r_0}A^{2/3}(2T),
\label{e:de_ld}
\end{eqnarray}
where $A=Z+N$ and the isospin is defined by $T=|Z-N|/2$. Hence, by comparing 
Eqs.~(\ref{e:diff}) and (\ref{e:de_ld}), it is seen that the $b$-coefficient 
is expected to increase as $A^{2/3}$. 

For a comparison with experimental $b$-coefficients, we turn to the more 
sophisticated liquid-drop parameterization of Ref.~\cite{r:Mol88}. Here
the form of the ``Coulomb'' energy will be outlined, and the Coulomb
energy difference between analog nuclei will be evaluated using the 
parameters defined in Ref.~\cite{r:Mol88}.
In macroscopic models, the ``Coulomb'' 
contribution to the binding energy is~\cite{r:Mol88}
\begin{equation}
E_{Coul}=c_1\frac{Z^2}{A^{1/3}}B_3 - c_4\frac{Z^{4/3}}{A^{1/3}}+
f(k_fr_p)\frac{Z^2}{A}-c_a(N-Z),
\label{e:mol_ld}
\end{equation}
where $c_1=3e^2/5r_0$ and $c_4=5/4(3/2\pi)^{2/3}c_1$.
The first two terms in Eq.~(\ref{e:mol_ld}) 
are the direct and exchange Coulomb energies,
the third is the proton-form factor correction, and the last is the
charge-asymmetry energy. 
The factor $B_3$ is the shape-dependent relative
Coulomb energy, which, to leading order for a spherical shape, is given by
\begin{equation}
B_3=1-\frac{5}{y_0^2}+\frac{75}{8y_0^3}-\frac{105}{8y_0^5},
\end{equation}
with $y_0=(r_0/a_{den})A^{1/3}\sim 1.657A^{1/3}$. The proton form factor
$f$ is dependent on the Fermi wave number $k_f=(9\pi Z/4A)^{1/3}(1/r_0)$
and the proton rms radius $r_p=0.8$~fm (see Eq.~(8) of Ref.~\cite{r:Mol88}), 
and for nuclei with $A\sim 50$ and $Z\sim A/2$ may be accurately approximated by 
$f=-0.214$~MeV. Using $r_0=1.16$~fm and $c_a=0.145$~MeV from
Ref.~\cite{r:Mol88}, the $b$-coefficient derived from the Coulomb-energy 
difference between analog nuclei is
\begin{equation}
b_{LD}=[0.7448A^{2/3}-1.882+1.535A^{-1/3}-0.7828A^{-1}]~{\rm MeV}.
\label{e:b_nix}
\end{equation}

Shown in Fig.~\ref{fig:fig1} is a comparison between Eq.~(\ref{e:b_nix}) 
(solid line) and experimental $b$-coefficients (solid squares). 
The experimental data comprise 116 $b$-coefficients, and were taken from 
Tables 3-7 in Ref.~\cite{r:Orm89} and the known ground-state analog mass 
differences tabulated in Ref.~\cite{r:Aud93} for $A \le 59$. 
In the figure, all the $b$-coefficients for a
given mass number were averaged together, and error bar reflects both the standard
deviation and the experimental uncertainties. Generally speaking, for a given
$A$, the $b$-coefficients are roughly constant, with the mean standard deviation
being 61~keV. From the microscopic point of view, deviations from constancy 
can be expected for two reasons. First, in some cases single-particle orbits from 
different major oscillator shells come into play, as in $15 \le A \le 17$, and 
second, near the limits of stability, the single-particle orbits are nearly
unbound, and the Thomas-Erhman shift~\cite{r:Tho52,r:Ehr51} 
needs to be accounted for (see Section~\ref{sec:Shell}). 

From Fig.~\ref{fig:fig1}, it is evident that the experimental $b$-coefficients
exhibit a global $A^{2/3}$ behavior. On the other hand, Eq.~(\ref{e:b_nix}) 
tends to underestimate the $b$-coefficients for $A < 40$, and the rms
deviation with the data is 138~keV. Within the context of a global 
parameterization, a slight improvement on Eq.~(\ref{e:b_nix}) can be
obtained by fitting to the experimental data, and an rms deviation of 102~keV
is achieved with 
\begin{equation}
b=[0.710A^{2/3}-0.946]~{\rm MeV},
\label{e:bfit}
\end{equation} 
which is also represented in Fig.~\ref{fig:fig1} by the dashed line.
For the most part, Eq.~(\ref{e:bfit}) leads to global description 
of the Coulomb energy differences between analog nuclei with an accuracy
of the order $100|Z-N|$~keV. To improve upon this, it is necessary to account 
for local nuclear structure via a microscopic model, which is the topic of 
the next section.

\section{Shell-Model Calculations of Coulomb Energy differences}
\label{sec:Shell}

In this section, the procedure for computing the Coulomb energy difference
between analog nuclei within the framework of the shell model is outlined. In
Ref.~\cite{r:Orm89}, empirical isovector and isotensor, or 
isospin-nonconserving (INC), Hamiltonians were determined for several shell-model
spaces by constraining them to reproduce experimental $b$- and $c$-coefficients. 
The primary components of the empirical interactions were the Coulomb
interaction and two-body isovector and isotensor interactions. 
In general, the empirical two-body isovector interaction was rather 
weak, while the isotensor interaction was found to   
to be consistent with the differences observed in the 
proton-proton and proton-neutron scattering lengths. 
The deviations between theoretical and experimental $b$ and $c$ coefficients
were of the order of 30~keV and 15~keV, respectively. 

In this work, proton-rich nuclei in the mass range $46 \le A \le 70$ are
investigated. For all but two cases (the $T=1/2$, $A=69$ isodoublet and the
$T=1$, $A=70$ isotriplet), 
the binding energy of the neutron-rich
analog has been measured and is tabulated in Ref.~\cite{r:Aud93}.
The shell-model calculations were performed using the shell-model code 
OXBASH~\cite{r:oxbash} in proton-neutron formalism using the  
configuration space defined by the $0f_{7/2}$, $0f_{5/2}$, 
$1p_{3/2}$, and $1p_{1/2}$ orbitals (the {\it fp} shell) and the  
FPD6 Hamiltonian given in Ref.~\cite{r:Ric90}. 
Here, instead of computing the $b$-coefficients, the ``Coulomb'' energy differences
between analogs were computed directly by adding the INC interaction to the
FPD6 Hamiltonian. Due to the large dimensions
present, some truncations on the model space were found to be necessary.
For $A\le 59$, all configurations contained within the $0f_{7/2}$ and 
$1p_{3/2}$ orbits were included, and the truncations were based on the
number of particles permitted to be excited out of the 
$0f_{7/2}$ and $1p_{3/2}$ orbits
into the $0f_{5/2}$ and $1p_{1/2}$ orbits. Generally, this ranged from
two to four particles so that the total dimensions (with good angular 
momentum) were less than 14,000. For $A > 60$, the $0f_{7/2}$
orbit was taken to be a closed core, and the $0f_{5/2}$, 
$1p_{3/2}$, and $1p_{1/2}$
single-particle energies of the FPD6 interaction were modified so as to
reproduce the levels of $^{57}$Ni under this assumption. 
For the most part, it was found that
the Coulomb energy shifts were not particularly sensitive to the 
applied truncations, and in a few cases where the effects of the 
truncations were tested, 
differences of only a few keV were found. Therefore, the
applied model-space truncations are not expected to provide a significant 
contribution to the uncertainty in the computed binding energies.  

The INC interaction used here consists of a Coulomb plus charge-dependent
(isotensor) interaction, and is described in detail in Ref.~\cite{r:Orm89}.
An important parameter for this interaction is the oscillator frequency,
$\hbar\omega$, since the Coulomb components are scaled as a function of 
$A$ by the factor~\cite{r:Orm89}
\begin{equation}
S(A) = \left[ \frac{\hbar\omega (A)}{11.096} \right]^{1/2},
\end{equation}
Generally, $\hbar\omega$ is chosen to reproduce experimental rms
charge radii, and for many nuclei  
it can be accurately parameterized by
\begin{equation}
\hbar\omega (A) = 45 A^{-1/3} - 25 A^{-2/3}~{\rm MeV}.
\label{e:hbw}
\end{equation}
It is important to note, however, that for $A \ge 45$, Eq.~(\ref{e:hbw})
underestimates $\hbar\omega$ as compared to values derived from experimental 
charge radii. Indeed, in Ref.~\cite{r:Orm89} the value of 10.222~MeV was used
for $A=53$ as opposed to the value of 10.208~MeV implied by
Eq.~(\ref{e:hbw}). In addition, the {\it fp}-shell INC interaction was refit 
in Ref.~\cite{r:Orm96b}, where it was found that  
better overall agreement between theoretical and experimental Coulomb
energy shifts was obtained using oscillator frequencies derived
from the rms charge radii of Hartree-Fock calculations  
using the Skyrme M$^*$ interaction. These values of $\hbar\omega$ are
tabulated in Table~\ref{tab:1}, and are used in the present work. 

In Refs.~\cite{r:Orm89,r:Orm96b}, the fitted INC interactions were able
to reproduce the experimental $b$-coefficients for {\it fp}-shell nuclei 
with an rms deviation of approximately 33~keV. 
However, the most difficult parameters to determine 
for the INC interaction are the Coulomb single-particle energies for the
$0f_{5/2}$ and $1p_{1/2}$ orbits, as there is very little experimental data
available that is sensitive to these quantities. In Ref.~\cite{r:Orm96b}, these 
single-particle energies were fit upon by making assumptions regarding spin
assignments for excited levels in $^{57}$Cu and $^{59}$Zn. In
retrospect, these levels are probably not appropriate for determining parameters 
for heavier nuclei because of uncertainties in 
spin assignments and the fact that the levels comprising the assumed 
doublet at 1.040~MeV in $^{57}$Cu~\cite{r:Bha92} are unbound, and 
strong Thomas-Ehrman shifts~\cite{r:Tho52,r:Ehr51} 
may apply (see below). Also, shell-model calculations for the 
$J^\pi=1/2^-$ and $5/2^-$ states in $^{59}$Zn indicate that these levels are
predominantly $1p_{3/2}^3$ configurations. On the other hand, the
the beta-endpoint energies for both $^{62}$Ga~\cite{r:Aud93} and 
$^{66}$As~\cite{r:Bha90} are 
sensitive to the $0f_{5/2}$ Coulomb single-particle energy 
and were used to help fix this parameter. In regards to the $1p_{1/2}$ 
single-particle energy, however, no data exists that will definitively set 
this parameter. For this reason, the value obtained in Ref.~\cite{r:Orm96b},
which also happens to reproduce the $b$-coefficients for the assumed $1/2^-$ 
states in $A=57$ and 59, is used.

Two additional concerns that affect this work are: (1) whether the 
{\it fp}-shell alone is sufficient to describe the nuclei in question, 
and (2) the effect of the 
Thomas-Ehrman shift on the Coulomb displacement energies near the drip line. 
As a measure of the appropriateness of just the {\it fp}-shell for the 
calculations, we
examine the the excitation energies of the first $J=2^+$ states in 
$N=Z$, even-even nuclei in the region $60 \le A \le 80$. Shown in 
Table~\ref{tab:2},
are the experimental~\cite{r:Fir96,r:Lis88} excitation energies 
of the these states in comparison with the
values obtained with the FPD6$^*$ interaction (FPD6 modified as indicated above after 
closing the $0f_{7/2}$ orbit). Overall, there is good agreement between the calculated 
and experimental values until $A = 76$, where there is a sudden drop in the 
excitation energy, which is an indication of the onset of collective behavior that
would necessitate the inclusion of orbits from the next major shell, such as the 
$0g_{9/2}$ orbit. Given the results in Table~\ref{tab:2}, the {\it fp}-shell 
is sufficient to describe the nuclei studied in this work.

In general, Coulomb energies are computed using harmonic oscillator, or sometimes 
bound Woods-Saxon, single-particle wave functions for the protons, with the length
scale chosen to reproduce experimental rms charge radii. Near the drip
line, however, this approximation can be inadequate. Because they are
loosely bound,
the proton single-particle wavefunctions are pushed out of the nuclear interior, and, 
as a consequence, the Coulomb energy is reduced. This shift in the Coulomb energy
was first noted by Thomas~\cite{r:Tho52} and Ehrman~\cite{r:Ehr51} in the 
$A=13$ system, and is most important for light nuclei where the
Coulomb barrier, which acts to confine the wave function in the nuclear interior,
is smaller, and for orbits with little or no centrifugal barrier, eg., the $s_{1/2}$ 
orbitals. This effect is well illustrated by the single-particle states 
in $A=17$, where the Coulomb 
displacement energy of the $J^\pi=5/2^+$ (the $0d_{5/2}$ orbit) ground state is 
3.543~MeV, while 
the shift for the $J^\pi=1/2^+$ state, which is a $1s_{1/2}$ single-particle state 
that is bound by only  107~keV, is 3.168~MeV. The influence of the centrifugal 
barrier is also apparent in these nuclei, as the Coulomb shift for the 
$J^\pi=3/2$ state (the $0d_{3/2}$ spin-orbit partner of the ground state),
which is unbound by 4.5~MeV, is 3.561~MeV.
On theoretical grounds, there are also self-consistent 
calculations~\cite{r:Naz96} that suggest that Thomas-Ehrman shifts for nuclei near  
the drip line may be as large a few hundred keV. Because of the empirical nature of the 
INC interaction, however, it is not clear how much of the Thomas-Ehrman effect has 
been absorbed into the interaction by fit. In addition, for nuclei near the drip-line,
the estimate of the theoretical uncertainty is of the order of a 100-250 ~keV, and
the effects of the Thomas-Ehrman shift are likely to lie within the 
quoted uncertainties for the absolute binding energy.

The parameters for the INC interaction used in this work are:
$\epsilon(0f_{7/2})=7.487$~MeV, 
$\epsilon(1p_{3/2})=7.312$~MeV,
$\epsilon(0f_{5/2})=7.337$~MeV, $\epsilon(1p_{1/2})=7.240$~MeV, $S_C=1.006$,
$S_0^{(1)}=0.0$, and $S_0^{(2)}=-4.2\times 10^{-2}$. Lastly, because of the
difficulties associated with determining the $0f_{5/2}$ and $1p_{3/2}$ 
single-particle energies, the uncertainties in the theoretical estimates
of the $b$-coefficients for nuclei with $A > 60$ are increased from
33~keV to 45~keV. 

Shown in Table~\ref{tab:3} are the results obtained for proton-rich nuclei 
whose binding energies are unknown in the mass region  
$46 \le A \le 69$. The table lists the experimental binding
energy of the neutron-rich analog, the predicted binding energy, one- and
two-proton separation energies, as well as the $Q$-value for electron
capture ($Q_{EC}$). The ground-state spins were taken from Ref.~\cite{r:Fir96} and
are also listed in the table. Wherever available, experimental binding
energies tabulated in Ref.~\cite{r:Aud93} were used in conjunction
with theoretical values to compute $Q_{EC}$ and the separation energies.

In addition to the nuclei listed in Table~\ref{tab:3}, predictions 
for the ``Coulomb'' energy differences for the the 
$T=1/2$, $A=69$ isodoublet and the $T=1$, $A=70$ isotriplet, for which the 
binding energy of the neutron-rich
member has not yet been measured experimentally, are given in
Table~\ref{tab:3a}. The theoretical uncertainties for $A=70$ include an
uncertainty of 20~keV in the {\it c} coefficient of Eq.~(\ref{e:IMME}).

Shown in Fig.~\ref{fig:fig2}, is a comparison of the binding energies reported here
with those from three other theoretical studies. This comparison is illustrated by the 
difference $\Delta BE= BE({\rm this~work})-BE({\rm other~work})$, which  
is plotted in the figure as a function of mass number, with the ordering the 
same as in Table~\ref{tab:3}. 
The error bars plotted at $\Delta BE=0$ represent the 
theoretical uncertainty of the binding energies listed in Table~\ref{tab:3}.
The open circles show the comparison with the previous shell-model 
calculations of Ormand in Ref.~\cite{r:Orm96} ($A\le 48$), the open 
triangles the comparison with the binding energies of Cole in 
Ref.~\cite{r:Col96} ($A\le 52$), 
while the solid squares represent the comparison with 
the binding energies obtained from the unified macroscopic-microscopic model
of M\"oller and Nix~\cite{r:Mol88} ($46 \le A \le 70$). While the results of 
Refs.~\cite{r:Orm96,r:Col96} are in agreement with those reported here, 
those of M\"oller and Nix are in severe
disagreement for some nuclei. The origin of these differences is two
fold. First, in the M\"oller-Nix study, the Coulomb-energy difference 
between analog nuclei is considerably smaller than in this work. This is 
illustrated in Fig.~\ref{fig:fig3}, where the $b$-coefficients for the
nuclei listed in Table~\ref{tab:3} (open triangles) are plotted as a 
function of $A$ in comparison with those derived from the M\"oller-Nix
masses (solid squares). In the mass
region $52 \le A \le 64$ the M\"oller-Nix $b$-coefficients are generally
100-200 keV smaller than those determined here. In addition, the
M\"oller-Nix $b$-coefficients are also in disagreement experimental trends,
as is evidenced by the $^{59}$Zn-$^{59}$Cu binding energy differences, 
where the M\"oller-Nix $b$-coefficient is 9.683~MeV, which is
200~keV smaller than the experimental value of 9.881(40)~MeV~\cite{r:Aud93}.
The second reason for the large disagreement in Fig.~\ref{fig:fig2} can be 
attributed to poor reproduction of the mass of the neutron-rich analog 
nucleus. For example, the M\"oller-Nix $^{59}$Cu mass excess is 
-54.8~MeV, which is in considerable disagreement with the experimental
value of -56.3515(17)~MeV tabulated in Ref.~\cite{r:Aud93}.

Also shown in Fig.~\ref{fig:fig3} is a comparison between the theoretical 
{\it b}-coefficients and the systematic trends expected from the 
liquid-drop model of Eq.~(\ref{e:b_nix}) (solid line) and the fit of 
Eq.~(\ref{e:bfit}) (dashed line). For the most part, the shell-model 
{\it b}-coefficients derived from the nuclei listed in Table~\ref{tab:3}
are in good agreement with the fitted parameterization of
Eq.~(\ref{e:bfit}), although they tend to be somewhat smaller than the 
systematic trend in the regions $A < 50$ and $A > 66$. Note that for 
$A < 50$, this is a continuation of the trend for experimental data 
as is observed in Fig.~\ref{fig:fig1} for $40 \le A \le 50$. 

As a further illustration of the systematic trend for the Coulomb energy
shifts, we examine the halflives for the Fermi transition between 
analog $J^\pi=0^+$, $T=1$ states in 
$N=Z$, odd-odd nuclei with $A=62$, 66, and 70. 
The partial half-life for the $\beta$ decay from the parent ground state 
to the $i^{th}$ state in daughter nucleus is given by
\begin{equation}
t_{1/2}^i = \frac{K}{G_V^2\mid {\cal M}_{o \rightarrow i} \mid^2 
f_{o \rightarrow i}},
\end{equation}
where $K=2\pi^3(\ln 2)\hbar^7/(m_e^5c^4)$ and 
$K/G_V^2=6170\pm 4 $~s~\cite{r:Wil78}. 
The statistical rate function, $f_{o \rightarrow i}$, 
depends on the beta end-point energy and here it is evaluated using
Eq.~(10) of Ref.~\cite{r:Orm96}. For a pure Fermi transition between
$T=1$ analog states, the transition matrix element 
${\cal M}_{o \rightarrow i}$ is given by $\sqrt{2(1-\delta_C)}$, where
$\delta_C$ is a small correction due to isospin-symmetry breaking.
Recent calculations~\cite{r:Orm96b} indicate that 
for these nuclei $\delta_C$ is expected to be of the order 1-2\%, and
for the purpose of comparing with experimental data, will be taken to
be equal to 1.5\%. In general, Gamow-Teller transitions to excited
states may also take place, and would tend to decrease the total beta-decay
halflife. However, not only are the matrix elements for these
transitions much smaller than for the Fermi transition, but since
they occur to states in the daughter nucleus at a higher excitation energy, 
the statistical rate function is also much smaller. Hence, to a good 
approximation, the beta decay of these nuclei may be taken to be pure Fermi. 
Listed in Table~\ref{tab:beta} are the 
predicted {\it Q}-values for electron capture as well as a comparison 
between the experimental~\cite{r:Bha90,r:Kin90,r:Bha93} and 
predicted beta-decay halflives. Given the
fact that the statistical rate function strongly depends on the beta
end-point energy (to the fifth power), the excellent agreement between
the experimental and predicted beta-decay halflives for all three nuclei 
is a good indication that the overall systematic behavior of the Coulomb 
energy differences is well reproduced here.   

This section is concluded with a discussion on $^{65}$As,
which is important from an astrophysical point of view. Because of the
long beta-decay halflife for $^{64}$Ge, if $^{65}$As were significantly
proton unbound, $^{64}$Ge would then become a ``waiting point'' in the 
{\it rp}-process and would inhibit the production of heavier elements. 
If, however, the halflife of $^{65}$As is dominated by beta decay, 
the {\it rp}-process will proceed through $^{65}$As primarily by proton 
capture to $^{66}$Se, although photodisintegration may begin to play
an important role if $^{65}$As is proton unbound~\cite{r:Cha92}.
 
From Table~\ref{tab:3}, $^{65}$As is found in this work to be unbound to
proton emission by 0.428(254)~MeV, with most of the uncertainty 
due to the uncertainty in the binding energy for $^{64}$Ge  
(0.250~MeV). On the other hand, $^{65}$As has been observed 
experimentally~\cite{r:Win93} with a beta-decay halflife of 
$190^{+110}_{-70}$~ms. From the fact that no protons were observed in the 
stopping detector during this experiment, it may be inferred that the partial
halflife for proton emission is significantly longer, and must be  
greater 1~s. The partial halflife for proton emission may be estimated
using the WKB approximation, which is outlined in some detail in the
next section [see Eq.~(\ref{e:WKB})], and, in particular for 
proton emission, in Ref.~\cite{r:Hof95}. A shell-model calculation within 
the {\it fp}-shell assuming a closed $0f_{7/2}$ orbit and the FPD6$^*$
interaction yields 0.13 for the spectroscopic factor $\theta^2$. 
Using the potential parameters of Ref.~\cite{r:Hof95}, a partial halflife
for proton emission longer than 1~s requires the one-proton separation
energy to be greater than -0.23~MeV, which is in
agreement with the value given in Table~\ref{tab:3}. 
Because of the extreme sensitivity on the separation energy, however, it 
may never be possible to give a reasonable prediction for the partial
halflife for proton emission without explicitly measuring the masses for
both $^{64}$Ge and $^{65}$As. With the present uncertainty of 0.254~MeV, a 
range of 16 orders of magnitude is found for the halflife, i.e., between 
$1.4\times 10^{-12}$ to $1.6\times 10^{4}$~s. On the other hand,
supposing that the binding energy of $^{64}$Ge could be measured to within
a few keV, a theoretical uncertainty of $\sim 50$~keV remains for
$^{65}$As, which for a separation energy of -0.2~MeV, leads to a range
of nearly four orders of magnitude in the proton partial halflife.

\section{Identification of candidates for di-proton emission}
\label{sec:diproton}

In this section, the partial half-lives for di-proton emission are examined, 
with the intention of identifying candidates amenable to 
experimental detection while taking into account theoretical uncertainties. 
As was mentioned in the introduction, the range of observable lifetimes
for di-proton emission 
is limited by competing decay mechanisms and 
experimental setups. In general, all candidates for di-proton emission
have large $\beta$-endpoints, and, as a consequence, the $\beta$-decay 
halflives are expected to be of the order 1-100~ms~\cite{r:Orm96}. Also,
in several experiments, such as in Ref.~\cite{r:Bla96}, the initial 
nucleus must live long enough to be identified. In this case, the
limiting time is determined by the time-of-flight in the experimental 
apparatus. In general, these two conditions impose a practical limit on
the observable halflife for di-proton emission to be in the range
$10^{-8}-10^{-3}$~s.

In Refs.~\cite{r:Bro91,r:Orm96}, the di-proton decay halflives were
estimated using $r$-matrix theory~\cite{r:rmat} 
while taking the channel radius, $R_0$, 
to be 4~fm for all cases. In contrast, in Ref.~\cite{r:Naz96} the halflife 
for $^{48}$Ni, was estimated using the 
Wentzel-Kramers-Brillouin (WKB) approximation. Because of
uncertainties associated with the choice of the channel radius, the 
WKB approximation for the di-proton decay halflife will be used here.
Following Ref.~\cite{r:Naz96}, the WKB expression for the partial 
decay width is
\begin{equation}
\Gamma_{2p}=\theta^2{\cal N}\frac{\hbar^2}{4\mu}
\exp\left[ -2\int_{r_{in}}^{r_{out}} dr k(r))\right],
\label{e:WKB}
\end{equation}
where $\theta^2$ is the spectroscopic factor for finding the di-proton
in the correlated $L=0$ state,
$\mu$ is the reduced mass, $r_{in}$ and $r_{out}$ are the classical
inner and outer turning points, respectively, the normalization factor
${\cal N}$ is determined by 
\begin{equation}
{\cal N}\int_0^{r_{in}} dr\frac{1}{k(r)}\cos^2
\left [ \int_0^r dr^\prime k(r^\prime )-\frac{\pi}{4} \right ] =1,
\end{equation}
and $k(r)$ is the wave number given by
\begin{equation}
k(r)=\sqrt{\frac{2\mu}{\hbar^2}\frac{m^*(r)}{m}|Q_{2p}-V_{2p}(r)|}.
\label{e:kwave}
\end{equation}
In Eq.~(\ref{e:kwave}), the asymptotic energy of the di-proton is 
$Q_{2p}=-S_{2p}$, $V_{2p}(r)$ is the average di-proton potential, and
$m^*(r)/m$ is the proton effective mass. As in Ref.~\cite{r:Naz96}, 
$V_{2p}(r)$ is approximated by $2V_p(r)$, where $V_p(r)$ is the 
self-consistent proton potential for the $(Z-2,N)$ nucleus obtained from
a Hartree-Fock or a Hartree-Fock-Bogoliubov calculation. Here, the 
halflives were computed using Hartree-Fock potentials using a 
Skyrme-type two-body interaction. It was found that the various Skyrme
interactions give halflives that are in agreement to within a factor of
two, and the results reported here were obtained using the Skyrme $M^*$ 
interaction. In addition, the halflives computed using Eq.~(\ref{e:WKB}) 
were found to be approximately an order of magnitude shorter than those
obtained using the $r$-matrix representation with $R_0=4$~fm (as was
used in Refs.~\cite{r:Bro91,r:Orm96}). On the other hand, if the channel
radius is chosen to be equal to the classical inner turning point, $r_{in}$,
the $r$-matrix approach yields halflives that are within a factor of two
of the WKB method.

The spectroscopic factor, $\theta$, can be evaluated within
the framework of the shell model. For di-proton emission the 
spectroscopic factor can be estimated using the 
cluster-overlap approximation~\cite{r:clus}, namely
\begin{equation}
\theta^2 = G^2[A/(A-k)]^\lambda |\langle \Psi_f | \psi_c | \Psi_i \rangle |^2,
\end{equation} 
where $k$, $\lambda$, and $G^2$ are parameters dependent on the model space
and the emitted cluster, and $\psi_c$ is a two-proton cluster wave function 
in which the relative motion of the particles is governed by the $0S$ state,
and is obtained by diagonalizing an SU3 conserving interaction within the
shell-model configuration space~\cite{r:clus}. 

Of all the quantities in Eq.~(\ref{e:WKB}), the di-proton decay rate is
most sensitive to the two-proton separation energy $S_{2p}$.
Indeed, it was illustrated in Ref.~\cite{r:Orm96} that an uncertainty of 
$\pm 100$~keV in a separation energy of the order 500~keV can lead to a
range of nearly six orders of magnitude in the di-proton decay halflife.
In contrast, the spectroscopic factors are expected to be of the order
0.5-0.75~\cite{r:Bro91}, and shouldn't lead to any more than an order
of magnitude decrease in the decay rate (increase in the halflife).
Given that the theoretical uncertainties in the separation energy 
for each of the di-proton emitters considered in this work are all greater 
than 175~keV, an accurate estimate of the spectroscopic factor 
is not needed in order to obtain an order-of-magnitude estimate of the 
di-proton halflife for the purpose of 
identifying the best candidates for experimental observation. 
Hence, the lifetimes reported here are evaluated
assuming $\theta^2=1$ with the understanding that they are probably 
too short by a factor of two to four.

Listed in Table~\ref{tab:4} are the halflives 
($t_{1/2}=\hbar\ln 2/\Gamma_{2p}$) 
associated with di-proton emission for all nuclei in Table~\ref{tab:3}
that are predicted to be unstable to two-proton emission while being 
bound to proton emission. Also, for the purpose of comparison, the halflives
for $^{38}$Ti, $^{45}$Fe, and $^{48}$Ni given in Ref.~\cite{r:Orm96} are
also listed in the table. Given the practical limitations on the
halflife for the experimental observation of this decay mode, the best
candidates are $^{45}$Fe, $^{48}$Ni, and $^{63}$Se. Of these three, 
perhaps the best is $^{45}$Fe since it is likely that it has already been
identified experimentally~\cite{r:Bla96}. On the other hand, 
both $^{59}$Ge and $^{67}$Kr have halflives that are long enough to make them
marginal candidates for experimental observation. 

\section{Conclusions}
\label{sec:Conclude}

In this work, Coulomb energy differences between mirror nuclei with 
$46 \le A \le 70$ were computed within the framework of the nuclear
shell model using an effective Coulomb plus isotensor interaction.
Absolute binding energies for proton-rich nuclei are predicted by adding
the Coulomb energy differences to the experimental binding energy of the
neutron-rich analog. With these binding energies, proton separation
energies are computed, and the location of the proton drip-line is
delineated. 

The computed Coulomb energy differences were also compared 
with systematic trends predicted by the liquid-drop model and a fit
to experimental {\it b}-coefficients assuming a $A^{2/3}$ dependence. It
was found that the shell-model calculations were in good agreement with
the systematic trends, except for $A \le 50$ and $A\ge 66$. As a further
test on the systematic trend of the shell-model Coulomb shifts,
halflives for the Fermi transitions in odd-odd, $N=Z$ nuclei with
$A=62$, 66, and 70 were computed and found to be in excellent agreement
with experimental data. The shell-model binding energies predicted here were 
also compared with three previous works. While the results of
Ormand~\cite{r:Orm96} (only for $A\le 48$) and 
Cole~\cite{r:Col96} (only for $A\le 52$) are in good agreement with those
reported here, those of M\"oller and Nix~\cite{r:Mol88} are not. It was
found that the disagreement with the M\"oller-Nix masses is due to
differences in both the Coulomb energy shifts and the binding
energy of the neutron-rich analog. For the most part, the data
presented in Fig.~\ref{fig:fig1} is the only data that is explicitly 
sensitive to a parameterization of the Coulomb energy. Given
the importance of analog symmetry and the overall success of the IMME, 
any global parameterization of binding energies should include a proper
description of the Coulomb energy differences. Towards this end, perhaps
the best approach is to determine the parameters of a microscopic-macroscopic 
model using the neutron-rich binding energies, while fixing 
the parameters of the Coulomb plus isovector part so as to reproduce 
the Coulomb energy shifts between mirror nuclei. Even in this limit,
however, it has to be noted that the systematic parameterization is
capable of reproducing the experimental {\it b} coefficients of the 
IMME only at the level of approximately 100~keV.

Finally, two-proton separation energies were also computed, and 
halflives associated with correlated di-proton emission were computed
using the WKB approximation. Given practical constraints on the halflife
for the observation of this decay mode imposed by competition with 
beta decay and experimental setups, the best candidates for experimental
observation are predicted to be $^{45}$Fe, $^{48}$Ni, and $^{63}$Se.

\begin{center}
{\bf Acknowledgments}
\end{center}
Discussions with B.A.~Brown, W.~Nazarewicz, and M.~Thoennessen are gratefully
acknowledged.
Oak Ridge National Laboratory is managed for the U.S. Department of Energy
by Lockheed Martin Energy Research Corp. under contract No.
DE--AC05--96OR22464. This work was supported in part by NSF Cooperative
agreement No. EPS~9550481, NSF Grant No. 9603006, and DOE contract
DE--FG02--96ER40985. 

\newpage
\bibliographystyle{try}

\begin{thebibliography}{11}
\bibitem{r:foot0} Assuming only two-body interactions, the IMME is exact at
the level of first-order perturbation theory. 
At present, the only known exception to the IMME is the
$J^\pi=3/2^-, T=3/2$ multiplet for $A=9$~\cite{r:Ben79}.
\end{thebibliography}

\newpage
\begin{table}
\caption{Values of $\hbar\omega$ used for {\it fp}-shell nuclei}
\begin{tabular}{cccc}
A & $\hbar\omega$ (MeV) & A & $\hbar\omega$ (MeV) \\
\tableline
 40 & 10.603 &  60 & 10.156  \\
 41 & 10.603 &  61 & 10.087 \\
 42 & 10.603 &  62 & 10.017 \\
 43 & 10.608 &  63 & 9.954 \\
 44 & 10.614 &  64 & 9.890 \\
 45 & 10.603 &  65 & 9.786 \\
 46 & 10.592 &  66 & 9.681 \\
 47 & 10.581 &  67 & 9.589 \\
 48 & 10.570 &  68 & 9.496 \\
 49 & 10.560 &  69 & 9.460 \\
 50 & 10.550 &  70 & 9.424 \\
 51 & 10.539 &  72 & 9.331 \\
 52 & 10.528 &  73 & 9.168 \\
 53 & 10.507 &  74 & 9.203 \\
 54 & 10.486 &  76 & 9.032 \\
 55 & 10.470 &  77 & 9.100 \\
 56 & 10.454 &  78 & 8.923 \\
 57 & 10.376 &  79 & 9.869 \\
 58 & 10.298 &  80 & 8.816 \\
 59 & 10.227 &  & \\
\end{tabular}
\label{tab:1}
\end{table}
\newpage

\begin{table}
\caption{Comparison between theoretical (with FPD6$^*$ and experimental excitation 
energies  (in MeV) of the first $J^\pi=2^+$ state in even-even $N=Z$ 
{\it fp}-shell nuclei}
\begin{tabular} {ccc}
$^AZ$ &  Expt. & FPD6$^*$ \\
\tableline
$^{60}$Zn  &  1.004$^a$  & 0.825  \\
$^{64}$Ge  &  0.902$^b$  & 0.700  \\
$^{68}$Se  &  0.854$^b$  & 0.600  \\
$^{72}$Kr  &  0.709$^b$  & 0.707  \\
$^{76}$Sr  &  0.261$^b$  & 0.752  \\
$^{80}$Zr  &  0.289$^b$  &   -    \\
\end{tabular}
\label{tab:2}
\end{table}
\noindent $^a$from Ref.~\cite{r:Fir96}\\
\noindent $^b$from ref.~\cite{r:Lis88}
\newpage

\begin{table}
\caption{Predicted binding energies, one- and two-proton separation energies
($S_p$ and $S_{2p}$, respectively), 
$\beta$-decay end-point energies for proton rich nuclei with $46 \le A \le
70$. The absolute binding energies were computed with theoretical Coulomb energy
shifts added onto the experimental binding energy for the neutron-rich analog,
also listed in the table.}
\begin{tabular}{ccccccccc}
  &  &  & $BE_{thy}$ &  & $BE_{exp}^{analog}$ &
$S_p$ & $S_{2p}$ & $Q_{EC}$ \\
$^AZ$ & $T_z$ & $J^\pi$ & (MeV) & $^AZ$-analog& (MeV) &
(MeV) & (MeV) & (MeV) \\
\tableline
$^{46}$Mn & 2 &  4$^+$  & 364.186(132) &  $^{46}$Sc & 396.610(1)   &   0.156(146) &   3.234(139) &  17.007(134) \\
$^{46}$Fe & 3 &  0$^+$  & 350.144(198) &  $^{46}$Ca & 398.769(2)   &   1.408(224) &   0.328(215) &  13.260(238) \\
$^{47}$Mn &3/2& 5/2$^-$ & 382.326(99)  &  $^{47}$Ti & 407.072(1)   &   0.351(101) &   5.237(100) &  12.020(100) \\
$^{47}$Fe &5/2& 7/2$^-$ & 365.973(165) &  $^{47}$Sc & 407.254(2)   &   1.787(211) &   1.943(177) &  15.571(192) \\
$^{47}$Co &7/2& 7/2$^-$ & 348.349(231) &  $^{47}$Ca & 406.045(2)   &  -1.795(304) &  -0.387(254) &     --       \\   
$^{48}$Mn & 1 &  4$^+$  & 397.101(66)  &  $^{48}$V  & 413.904(3)   &   1.973(68)  &   6.740(66)  &  13.579(66)  \\ 
$^{48}$Fe & 2 &  0$^+$  & 385.106(132) &  $^{48}$Ti & 418.698(1)   &   2.780(165) &   3.131(134) &  11.213(148) \\ 
$^{48}$Co & 3 &  6$^+$  & 365.153(198) &  $^{48}$Sc & 415.487(5)   &  -0.820(258) &   0.967(238) &     --       \\   
$^{48}$Ni & 4 &  0$^+$  & 348.854(264) &  $^{48}$Ca & 415.991(4)   &   0.505(351) &  -1.290(330) &  15.517(330) \\ 
$^{49}$Fe &3/2& 7/2$^-$ & 399.802(99)  &  $^{49}$V  & 425.457(1)   &   2.701(119) &   4.674(100) &  12.963(102) \\ 
$^{49}$Co &5/2& 7/2$^-$ & 384.184(165) &  $^{49}$Ti & 426.841(1)   &  -0.922(211) &   1.858(192) &      --      \\    
$^{49}$Ni &7/2& 7/2$^-$ & 365.830(231) &  $^{49}$Sc & 425.618(4)   &   0.677(304) &  -0.143(284) &  17.572(284) \\ 
$^{49}$Cu &9/2& 3/2$^-$ & 344.413(297) &  $^{49}$Ca & 421.138(4)   &  -4.441(397) &  -3.936(376) &     --       \\    
$^{50}$Co & 2 &  6$^+$  & 400.060(132) &  $^{50}$V  & 434.790(1)   &   0.258(165) &   2.959(148) &  16.585(145) \\
$^{50}$Ni & 3 &  0$^+$  & 385.693(198) &  $^{50}$Ti & 437.780(1)   &   1.509(258) &   0.587(238) &  13.585(238) \\ 
$^{50}$Cu & 4 &  5$^+$  & 362.299(264) &  $^{50}$Sc & 431.674(16)  &  -3.531(351) &  -2.854(330) &     --       \\   
$^{50}$Zn & 5 &  0$^+$  & 340.823(330) &  $^{50}$Ca & 427.491(9)   &  -3.590(444) &  -8.031(423) &     --       \\   
$^{51}$Co &3/2& 7/2$^-$ & 417.864(99)  &  $^{51}$Cr & 444.306(1)   &   0.164(116) &   4.317(102) &  12.868(100) \\ 
$^{51}$Ni &5/2& 7/2$^-$ & 401.684(165) &  $^{51}$V  & 445.841(1)   &   1.624(211) &   1.882(192) &  15.398(192) \\ 
$^{51}$Cu &7/2& 3/2$^-$ & 382.472(231) &  $^{51}$Ti & 444.153(1)   &  -3.221(304) &  -1.712(284) &     --       \\    
$^{52}$Co & 1 &  6$^+$  & 432.912(66)  &  $^{52}$Mn & 450.851(2)   &   1.398(68)  &   6.283(66)  &  14.003(67)  \\ 
$^{52}$Ni & 2 &  0$^+$  & 420.478(132) &  $^{52}$Cr & 456.345(1)   &   2.614(165) &   2.778(145) &  11.652(148) \\ 
$^{52}$Cu & 3 &  3$^+$  & 399.399(198) &  $^{52}$V  & 453.152(1)   &  -2.285(258) &  -0.661(238) &     --       \\    
$^{52}$Zn & 4 &  0$^+$  & 380.321(264) &  $^{52}$Ti & 451.961(7)   &  -2.151(351) &  -5.372(330) &     --       \\   
$^{53}$Ni &3/2& 7/2$^-$ & 435.558(99)  &  $^{53}$Mn & 462.905(2)   &   2.646(119) &   4.044(100) &  12.956(101) \\ 
$^{53}$Cu &5/2& 3/2$^-$ & 418.835(165) &  $^{53}$Cr & 464.285(2)   &  -1.643(211) &   0.971(192) &     --       \\    
$^{53}$Zn &7/2& 7/2$^-$ & 397.948(231) &  $^{53}$V  & 461.631(3)   &  -1.451(304) &  -3.736(284) &     --       \\     
$^{54}$Cu & 2 &  3$^+$  & 434.906(132) &  $^{54}$Mn & 471.844(2)   &  -0.652(165) &   1.994(148) &     --       \\   
$^{54}$Zn & 3 &  0$^+$  & 418.605(198) &  $^{54}$Cr & 474.004(1)   &  -0.230(258) &  -1.873(238) &  15.519(238) \\ 
$^{54}$Ga & 4 &  3$^+$  & 393.891(264) &  $^{54}$V  & 467.744(15)  &  -4.057(351) &  -5.508(330) &     --       \\    
$^{55}$Cu &3/2& 3/2$^-$ & 452.997(99)  &  $^{55}$Fe & 481.057(1)   &  -0.153(111) &   3.701(101) &  13.568(100) \\ 
$^{55}$Zn &5/2& 5/2$^-$ & 435.071(165) &  $^{55}$Mn & 482.071(1)   &   0.165(211) &  -0.487(192) &  17.144(192) \\ 
$^{55}$Ga &7/2& 3/2$^-$ & 414.644(231) &  $^{55}$Cr & 480.250(1)   &  -3.961(304) &  -4.191(284) &     --       \\     
$^{56}$Cu & 1 &  4$^+$  & 467.899(66)  &  $^{56}$Co & 486.906(1)   &   0.552(67)  &   5.166(66)  &  15.307(67)  \\
$^{56}$Zn & 2 &  0$^+$  & 454.214(132) &  $^{56}$Fe & 492.254(1)   &   1.217(165) &   1.064(141) &  12.903(148) \\ 
$^{56}$Ga & 3 &  3$^+$  & 432.226(198) &  $^{56}$Mn & 489.315(1)   &  -2.845(258) &  -2.680(238) &     --       \\     
$^{56}$Ge & 4 &  0$^+$  & 412.381(264) &  $^{56}$Cr & 488.507(10)  &  -2.263(351) &  -6.224(330) &     --       \\    
$^{57}$Zn &3/2& 7/2$^-$ & 469.440(99)  &  $^{57}$Co & 498.282(1)   &   1.541(119) &   2.093(100) &  14.461(100) \\ 
$^{57}$Ga &5/2& 3/2$^-$ & 451.874(165) &  $^{57}$Fe & 499.885(1)   &  -2.340(211) &  -1.123(192) &     --       \\     
$^{57}$Ge &7/2& 5/2$^-$ & 430.634(231) &  $^{57}$Mn & 497.992(3)   &  -1.592(304) &  -4.437(284) &     --       \\    
$^{58}$Ga & 2 &  2$^+$  & 468.039(132) &  $^{58}$Co & 506.855(2)   &  -1.401(165) &   0.140(148) &     --       \\   
$^{58}$Ge & 3 &  0$^+$  & 451.578(198) &  $^{58}$Fe & 509.945(1)   &  -0.296(258) &  -2.636(238) &  15.679(238) \\
$^{58}$As & 4 &  3$^+$  & 426.697(266) &  $^{58}$Mn & 504.480(30)  &  -3.937(352) &  -5.529(331) &     --       \\    
$^{59}$Ga &3/2& 3/2$^-$ & 486.040(99)  &  $^{59}$Ni & 515.453(1)   &  -0.920(111) &   1.357(100) &     --       \\    
$^{59}$Ge &5/2& 7/2$^-$ & 468.097(165) &  $^{59}$Co & 517.308(1)   &   0.058(211) &  -1.343(192) &  17.161(192) \\ 
$^{59}$As &7/2& 3/2$^-$ & 447.648(231) &  $^{59}$Fe & 516.526(1)   &  -3.930(304) &  -4.226(284) &     --       \\    
$^{60}$Ga & 1 &  2$^+$  & 500.080(66)  &  $^{60}$Cu & 519.933(3)   &   0.080(77)  &   2.971(66)  &  14.130(67)  \\  
$^{60}$Ge & 2 &  0$^+$  & 487.127(132) &  $^{60}$Ni & 526.842(1)   &   1.087(165) &   0.167(141) &  12.171(148) \\  
$^{60}$As & 3 &  5$^+$  & 465.094(198) &  $^{60}$Co & 524.800(1)   &  -3.003(258) &  -2.945(238) &     --       \\     
$^{61}$Ga &1/2& 3/2$^-$ & 515.179(48)  &  $^{61}$Zn & 525.223(16)  &   0.187(49)  &   5.307(48)  &   9.262(50)  \\
$^{61}$Ge &3/2& 3/2$^-$ & 501.415(135) &  $^{61}$Cu & 531.642(2)   &   1.335(150) &   1.415(141) &  12.982(143) \\ 
$^{61}$As &5/2& 3/2$^-$ & 484.381(225) &  $^{61}$Ni & 534.595(1)   &  -2.746(261) &  -1.659(246) &     --       \\    
$^{62}$Ge & 1 &  0$^+$  & 517.720(91)  &  $^{62}$Zn & 538.119(10)  &   2.541(102) &   2.728(91)  &   9.664(95)  \\
$^{62}$As & 2 &  1$^+$  & 499.816(180) &  $^{62}$Cu & 540.529(4)   &  -1.599(225) &  -0.264(192) &     --       \\   
$^{62}$Se & 3 &  0$^+$  & 484.239(270) &  $^{62}$Ni & 545.259(1)   &  -0.142(351) &  -2.888(301) &  14.795(325) \\
$^{63}$Ge &1/2& 3/2$^-$ & 530.597(110) &  $^{63}$Ga & 540.930(100) &   2.431(113) &   5.374(111) &   9.551(148) \\
$^{63}$As &3/2& 3/2$^-$ & 516.321(135) &  $^{63}$Zn & 547.232(2)   &  -1.399(163) &   1.142(143) &     --       \\   
$^{63}$Se &5/2& 3/2$^-$ & 499.885(225) &  $^{63}$Cu & 551.382(1)   &   0.069(288) &  -1.530(262) &  15.654(262) \\
$^{64}$As & 1 &  0$^+$  & 530.315(90)  &  $^{64}$Ga & 551.147(4)   &  -0.282(142) &   2.149(94)  &  14.853(266) \\
$^{64}$Se & 2 &  0$^+$  & 517.411(180) &  $^{64}$Zn & 559.094(2)   &   1.090(225) &  -0.309(202) &  12.122(201) \\
$^{65}$As &1/2& 3/2$^-$ & 545.522(46)  &  $^{65}$Ge & 556.010(10)  &  -0.428(254) &   4.592(110) &     --       \\     
$^{65}$Se &3/2& 3/2$^-$ & 531.473(135) &  $^{65}$Ga & 563.036(2)   &   1.158(162) &   0.876(174) &  13.267(143) \\
$^{65}$Br &5/2& 1/2$^-$ & 514.580(225) &  $^{65}$Zn & 567.020(2)   &  -2.831(288) &  -1.741(262) &     --       \\    
$^{66}$Se & 1 &  0$^+$  & 548.091(95)  &  $^{66}$Ge & 569.290(30)  &   2.569(105) &   2.141(267) &  10.087(112) \\
$^{66}$Br & 2 &  0$^+$  & 529.780(180) &  $^{66}$Ga & 572.176(3)   &  -1.693(225) &  -0.535(201) &     --       \\     
$^{66}$Kr & 3 &  0$^+$  & 514.579(270) &  $^{66}$Zn & 578.133(2)   &  -0.001(351) &  -2.832(325) &  14.419(325) \\
$^{67}$Se &1/2& 5/2$^-$ & 560.882(110) &  $^{67}$As & 571.610(100) &   1.922(125) &   4.872(110) &     --       \\    
$^{67}$Br &3/2& 1/2$^-$ & 546.355(135) &  $^{67}$Ge & 578.398(5)   &  -1.736(165) &   0.833(143) &     --       \\    
$^{67}$Kr &5/2& 1/2$^-$ & 529.935(225) &  $^{67}$Ga & 583.403(2)   &   0.155(288) &  -1.538(262) &  15.638(262) \\
$^{68}$Br & 1 &  3$^+$  & 560.365(135) &  $^{68}$As & 581.910(100) &  -0.517(174) &   1.405(147) &     --       \\    
$^{68}$Kr & 2 &  0$^+$  & 547.668(180) &  $^{68}$Ge & 590.792(6)   &   1.313(225) &  -0.423(204) &  11.915(225) \\
$^{68}$Rb & 3 &  1$^+$  & 526.980(270) &  $^{68}$Ga & 591.680(2)   &  -2.955(351) &  -2.800(325) &     --       \\    
$^{69}$Br &1/2& 3/2$^-$ & 575.737(54)  &  $^{69}$Se & 586.620(30)  &  -0.663(305) &   4.127(114) &     --       \\   
$^{69}$Kr &3/2& 5/2$^-$ & 561.477(138) &  $^{69}$As & 594.180(30)  &   0.075(193) &  -0.442(176) &  14.515(148) \\ 
$^{70}$Rb & 2 &  4$^+$  & 559.398(187) &  $^{70}$As & 603.520(50)  &  -1.042(232) &  -0.967(230) &     --       \\   
\end{tabular}
\label{tab:3}
\end{table}
\newpage

\begin{table}
\caption{Predictions for the ``Coulomb'' energy difference for the 
$T=1/2$, $A=69$ and $T=1$, $A=70$ nuclei.}
\begin{tabular}{ccc}
$^ZA-^{Z-1}A$ & $J^\pi$ & $\Delta BE$ (MeV) \\
\tableline
$^{69}$Br-$^{69}$Se & $3/2^-$ & -10.883(45)\\
$^{70}$Kr-$^{70}$Br & $0^+$   & -11.241(50)\\
$^{70}$Br-$^{70}$Se & $0^+$   & -10.801(50)\\
\end{tabular}
\label{tab:3a}
\end{table}

\begin{table}
\caption{Comparison between experimental and predicted beta-decay
halflives for odd-odd, $N=Z$ Fermi transitions. The predicted 
$Q_{EC}$-value is also given.}
\begin{tabular}{cccc}
$^ZA$ & $Q_{EC}$~(MeV) & $t_{1/2}$ (ms) & $t_{1/2}^{exp}$ (ms) \\
\tableline
$^{62}$Ga & 9.191(50) & 115(2) & 116.1(2)$^a$\\
$^{66}$As & 9.592(50) &  94(2) & 95.8(2)$^b$\\
$^{70}$Br & 10.019(50) & 76(3) & 79.1(8)$^c$\\
\end{tabular}
\label{tab:beta}
\end{table}
\noindent $^a$ From Ref.~\cite{r:Kin90}.\\
\noindent $^b$ From Ref.~\cite{r:Bha90}.\\
\noindent $^c$ From Ref.~\cite{r:Bha93}.\\

\newpage
\begin{table}
\caption{Half-lives for di-proton emitter candidates. Also listed are
the theoretical predictions for the one- and two-proton
separation energies}
\begin{tabular}{cccccc}
$^A$Z & $S_p$ (MeV) & $S_{2p}$ (MeV) & $t_{1/2}$ (s) 
& $t_{1/2}^{min}$ (s) & $t_{1/2}^{max}$ (s) \\
\tableline
          &             &   Ref. [7]   &    &      \\
\tableline
$^{38}$Ti &  0.438(164) & -2.432(132) & $9\times 10^{-16}$  
&  $4\times 10^{-16}$ &   $2\times 10^{-15}$  \\
$^{45}$Fe & -0.010(198) & -1.279(181) & $10^{-6}$ 
&  $10^{-8}$ &   $10^{-4}$  \\
$^{48}$Ni &  0.502(164) & -1.137(210) & $3\times 10^{-3}$ 
&  $10^{-5}$ &  4  \\
\tableline
          &             & This work   &    &      \\
\tableline
$^{48}$Ni &  0.505(351) & -1.290(330) & $4\times 10^{-6}$
&  $5\times 10^{-9}$ &   $0.09$  \\
$^{49}$Ni &  0.677(304) & -0.143(284) & $3\times 10^{49}$  
&  $6\times 10^{14}$ &   $\infty$  \\
$^{55}$Zn &  0.165(211) & -0.487(192) & $5\times 10^{14}$ 
&  $2\times 10^{9}$ &   $3\times 10^{30}$  \\
$^{59}$Ge &  0.058(211) & -1.343(192) & $10^{-3}$
&  $10^{-5}$ &   $0.3$  \\
$^{63}$Se &  0.069(288) & -1.530(262) & $6\times 10^{-5}$
&  $3\times 10^{-7}$ &   $5\times 10^{-2}$  \\
$^{64}$Se &  1.090(225) & -0.309(202) & $5\times 10^{32}$
&  $6\times 10^{17}$ &   $4\times 10^{79}$  \\
$^{66}$Kr & -0.001(351) & -2.832(325) & $3\times 10^{-12}$
&  $2\times 10^{-13}$ &   $6\times 10^{-11}$  \\
$^{67}$Kr &  0.155(288) & -1.538(262) & $2\times 10^{-3}$
&  $10^{-5}$ &   $0.2$  \\
$^{68}$Kr &  1.313(225) & -0.423(204) & $3\times 10^{24}$
&  $8\times 10^{13}$ &   $5\times 10^{49}$  \\
$^{69}$Kr &  0.075(193) & -0.442(176) & $2\times 10^{23}$
&  $2\times 10^{14}$ &   $10^{40}$  \\
\end{tabular}
\label{tab:4}
\end{table}
\newpage
\begin{figure}
\caption{Dependence of $b$-coefficients as a function of mass number $A$.
Experimental data are represented by the solid squares, while the 
values from the liquid-drop formula and the fit 
[Eqs.~(\ref{e:b_nix}) and (\ref{e:bfit})] are represented by the solid
and dashed lines, respectively.}
\label{fig:fig1}
\end{figure}

\begin{figure}
\caption{Difference between absolute binding energies listed in 
Table~\ref{tab:3} with those of M\"oller and Nix~[5] (solid squares),
Ormand~[7], and Cole~[8] (open triangles). The error bars at
$\Delta BE=0$ denote the theoretical uncertainty of the binding energies
listed in Table~\ref{tab:3}.} 
\label{fig:fig2}
\end{figure}

\begin{figure}
\caption{Dependence of theoretical {\it b}-coefficients for nuclei listed 
in Table~\ref{tab:3} (open triangles) as a function of mass number $A$. 
For comparison, the {\it b}-coefficients derived from the 
unified microscopic-macroscopic model of M\"oller and Nix~[5] 
are also shown (solid squares). The systematic behavior as expected 
from Eqs.~(\ref{e:b_nix}) and (\ref{e:bfit}) are represented by the solid and
dashed lines, respectively.}
\label{fig:fig3}
\end{figure}

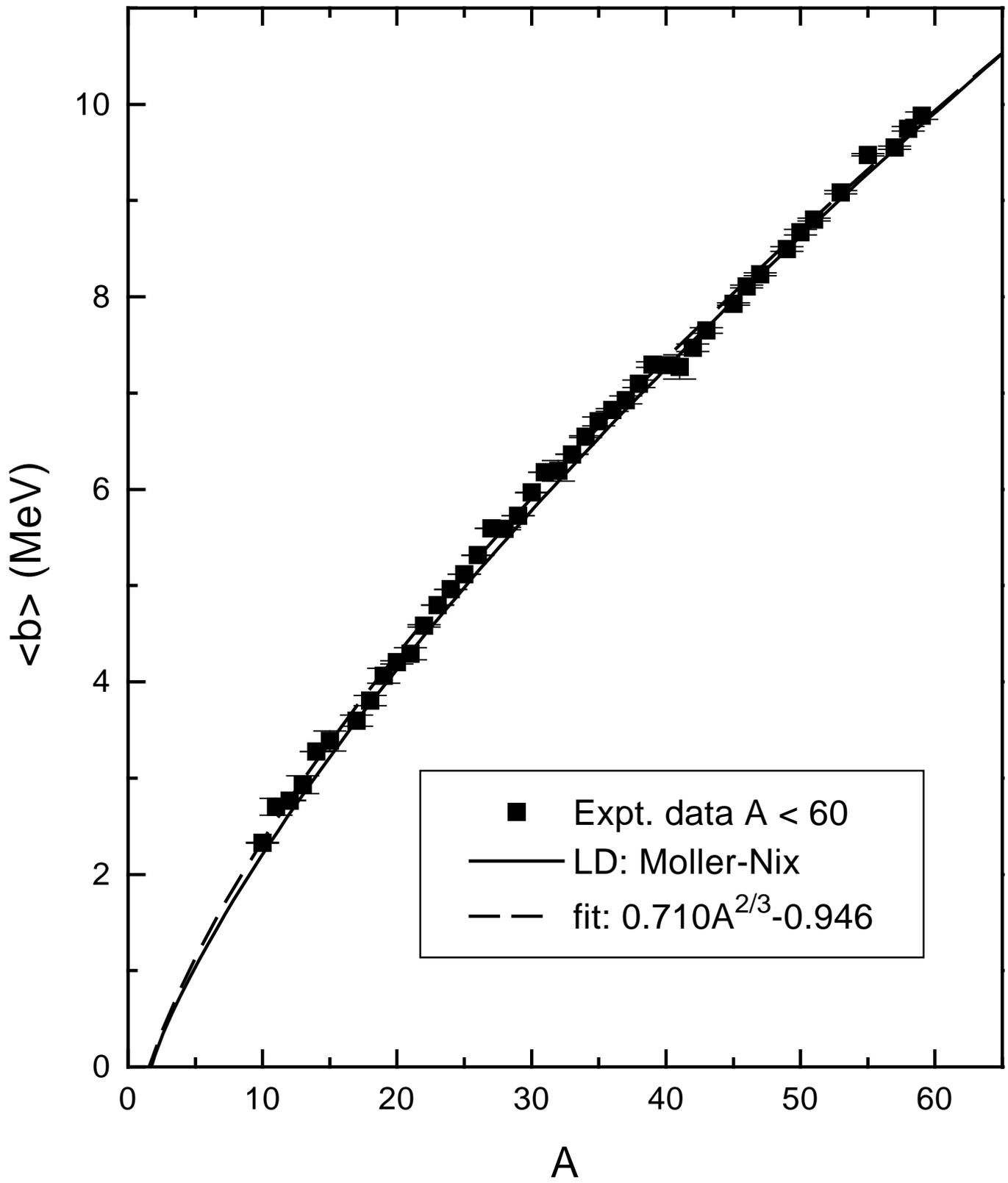
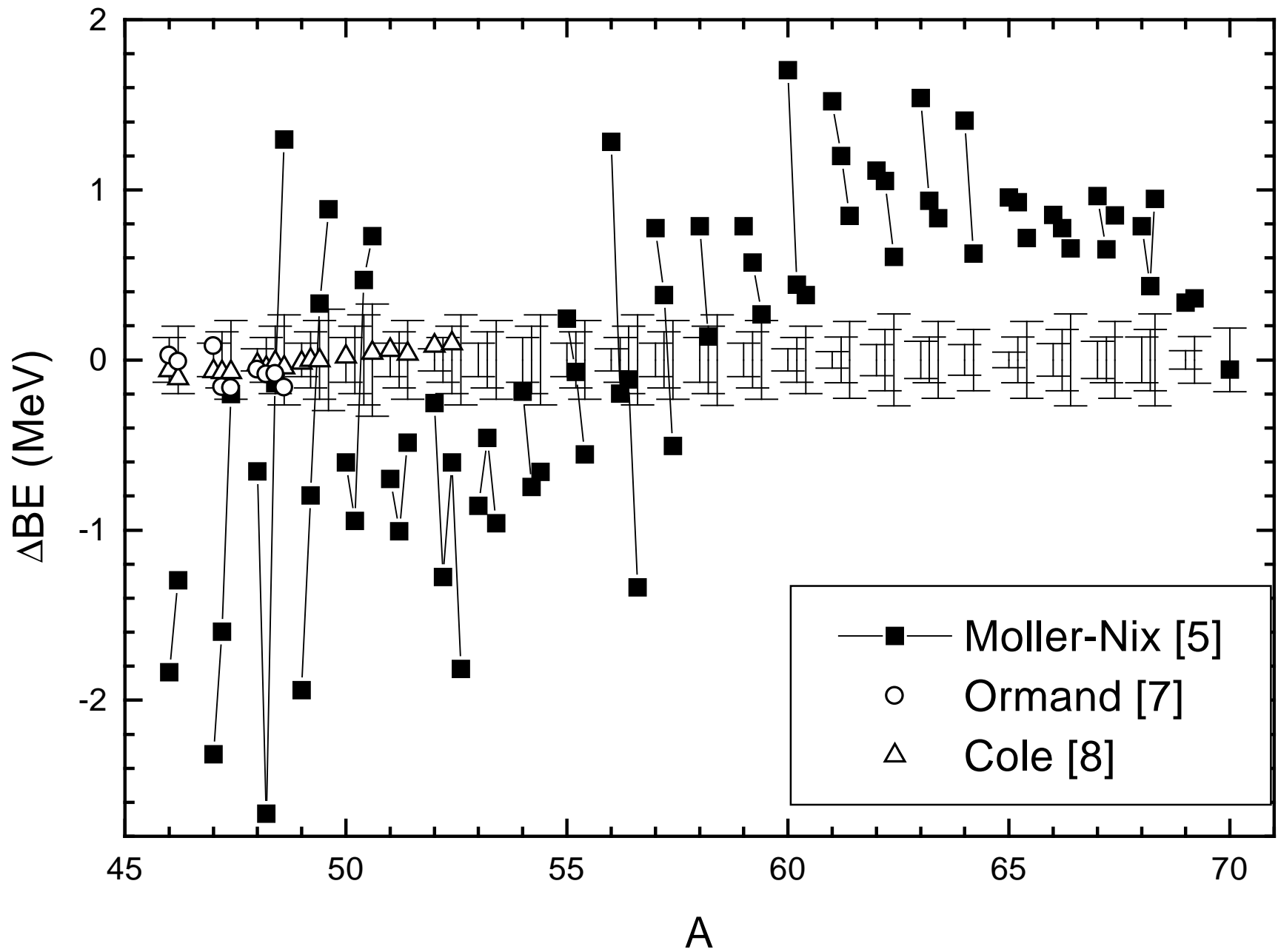
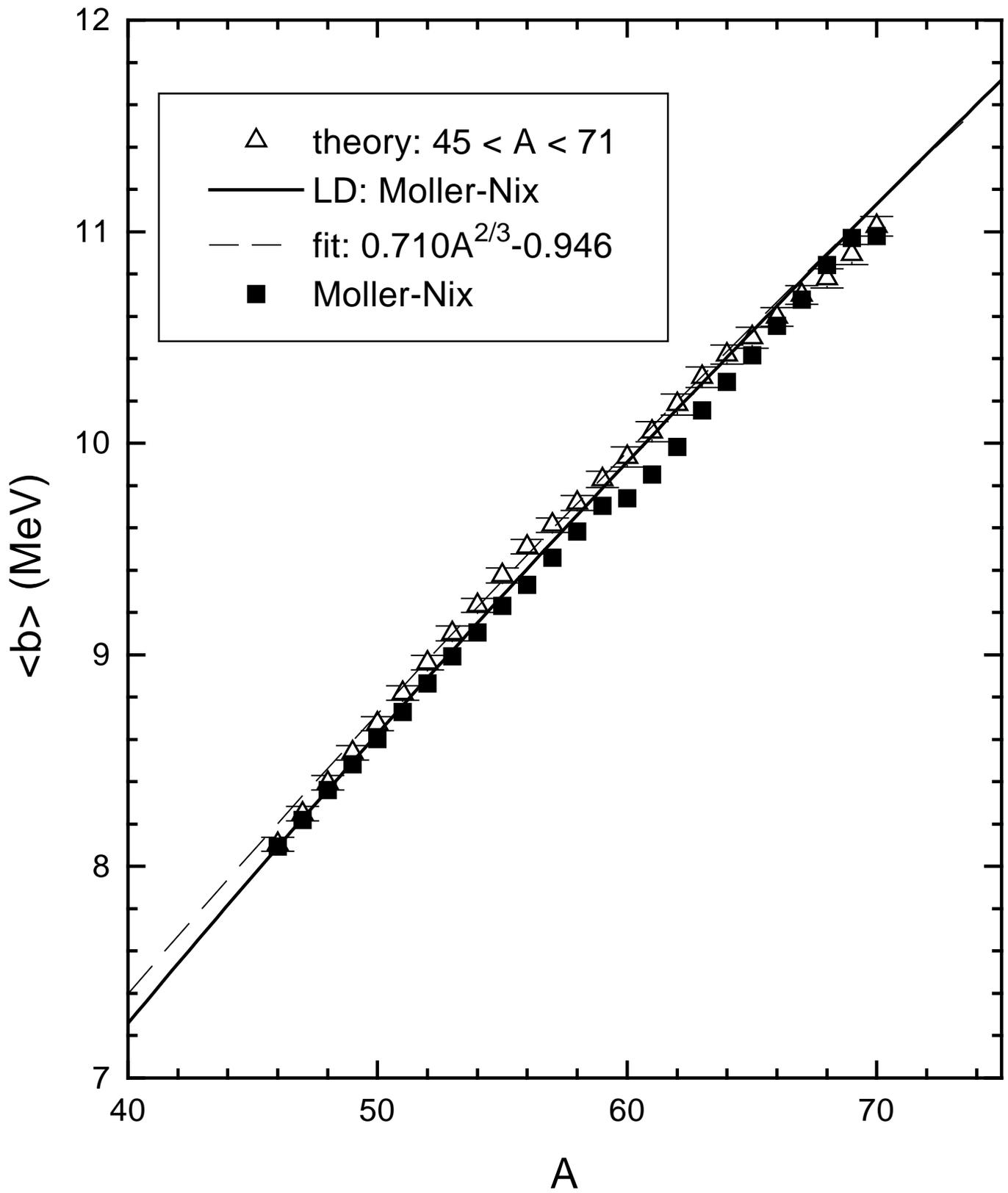
\end{document}